\documentclass[12pt,preprint]{emulateapj}

\usepackage{epsf}
\usepackage{color}
\usepackage{amsmath}
\usepackage{apjfonts}

\shorttitle{PHYSICAL PROPERTIES OF THE EIGHT O'CLOCK ARC}
\shortauthors{FINKELSTEIN ET AL.}

\newcommand{\sol}{$_{\odot}$}

\newcommand{\sfr}{M\sol~yr$^{-1}$}
\newcommand{\authorspace}{\vspace{-10pt}}
\def\arcs{\hbox{$^{\prime\prime}$}}

\begin{document}
\slugcomment{Accepted for Publication in the Astrophysical Journal}
\title{TURNING BACK THE CLOCK: INFERRING THE HISTORY OF THE EIGHT O'CLOCK ARC\altaffilmark{1}}

\author{\sc Steven L. Finkelstein\altaffilmark{2}}
\affil{George P. and Cynthia W. Mitchell Institute for Fundamental Physics and Astronomy, and\\ Department of Physics, Texas A\&M University, College Station, TX 77843-4242}
\author{\sc \authorspace Casey Papovich}
\affil{George P. and Cynthia W. Mitchell Institute for Fundamental Physics and Astronomy, and\\ Department of Physics, Texas A\&M University, College Station, TX 77843-4242}
\author{\sc \authorspace Gregory Rudnick}
\affil{Department of Physics and Astronomy, University of Kansas, Lawrence, KS 66045}
\author{\sc \authorspace Eiichi Egami}
\affil{Department of Astronomy, University of Arizona, Tucson, AZ 85721}
\author{\sc \authorspace Emeric Le Floc'h\altaffilmark{3}}
\affil{Institute for Astronomy, University of Hawaii, Honolulu, HI 96822}
\author{\sc \authorspace Marcia J. Rieke}
\affil{Department of Astronomy, University of Arizona, Tucson, AZ 85721}
\author{\sc \authorspace  Jane R. Rigby\altaffilmark{3}}
\affil{The Observatories of the Carnegie Institution of Washington, Pasadena, CA 91101}
\and
\author{\sc Christopher N. A. Willmer}
\affil{Department of Astronomy, University of Arizona, Tucson, AZ 85721}
\altaffiltext{1}{Based partly on observations obtained at the Gemini Observatory, which is operated by the Association of Universities for Research in Astronomy, Inc., under a cooperative agreement with the NSF on behalf of the Gemini partnership: the National Science Foundation (United States), the Science and Technology Facilities Council (United Kingdom), the National Research Council (Canada), CONICYT (Chile), the Australian Research Council (Australia), Ministerio da Ciencia e Tecnologia (Brazil) and SECYT (Argentina).  Some of the data presented herein were obtained at the W.M. Keck Observatory, which is operated as a scientific partnership among the California Institute of Technology, the University of California and the National Aeronautics and Space Administration. The Observatory was made possible by the generous financial support of the W.M. Keck Foundation.}
\altaffiltext{2}{Email: stevenf@physics.tamu.edu}
\altaffiltext{3}{Spitzer Fellow}

\begin{abstract}
We present the results from an optical and near-infrared spectroscopic study of the ultraviolet-luminous z = 2.73 galaxy, the 8 o'clock arc.  Due to gravitational lensing, this galaxy is magnified by a factor of $\mu$ $>$ 10, allowing in-depth measurements which are usually unfeasible at such redshifts.  In the optical spectra, we measured the systemic redshift of the galaxy, z = 2.7322 $\pm$ 0.0012, using stellar photospheric lines.  This differs from the redshift of absorption lines in the interstellar medium (ISM), z = 2.7302 $\pm$ 0.0006, implying gas outflows on the order of 160 km s$^{-1}$.  With H and K-band near-infrared spectra, we have measured nebular emission lines of H$\alpha$, H$\beta$, H$\gamma$, [N\,{\sc ii}] and [O\,{\sc iii}], which have a redshift z = 2.7333 $\pm$ 0.0001, consistent with the derived systemic redshift.  From the Balmer decrement, we measured the dust extinction in this galaxy to be A$_{5500}$ = 1.17 $\pm$ 0.36 mag.  Correcting the H$\alpha$ line-flux for dust extinction as well as the assumed lensing factor, we measure a star-formation rate of $\sim$ 270 \sfr,  which is higher than $\sim$ 85\% of star-forming galaxies at z $\sim$ 2--3.  Using combinations of all detected emission lines, we find that the 8 o'clock arc has a gas-phase metallicity of $\sim$ 0.8 Z\sol, showing that enrichment at high-redshift is not rare, even in blue, star-forming galaxies.  Studying spectra from two of the arc components separately, we find that one component dominates both the dust extinction and star-formation rate, although the metallicities between the two components are similar.   We derive the mass via stellar population modeling, and find that the arc has a total stellar mass of $\sim$ 4.2 $\times$ 10$^{11}$ M\sol, which falls on the mass-metallicity relation at z $\sim$ 2.  Finally, we estimate the total gas mass, and find it to be only $\sim$ 12\% of the stellar mass, implying that the 8 o'clock arc is likely nearing the end of a starburst.
\end{abstract}


\section{Introduction}

Although we now have the ability to discover galaxies as distant as z $\sim$ 7 (Iye et al.\ 2006), we do not yet have the technology available for detailed studies of these most distant of objects.  Rest-frame optical emission lines are excellent galactic diagnostics, but many of these are hard to detect at z $\gtrsim$ 3, due mainly to observational limitations.  One of the strongest diagnostic lines is H$\alpha$ ($\lambda$6563), which is shifted redward of the K-band at these redshifts.  At z $<$ 3, H$\alpha$ is observable in the near-infrared (NIR), but the sky background is high and these distant galaxies are intrinsically faint.  Thus, the observing time required to obtain scientifically useful data for these galaxies can be exorbitant, even on large telescopes.

Gravitationally lensed galaxies represent a prime opportunity for detailed studies at redshifts which are otherwise beyond reach.  The lensing foreground object smears out the light from the lensed background galaxy over a larger angular area, but the surface brightness is conserved, thus the galaxy has a brighter apparent magnitude.  The canonical example is the strongly lensed galaxy MS 1512--cB58 at z = 2.727 (Yee et al.\ 1996; hereafter cB58), which at the time of discovery was the brightest known Lyman break galaxy (LBG; V = 20.5).  While it was initially thought to be the most active star-forming galaxy with a star formation rate (SFR) $\sim$ 10$^{3}$ \sfr, Williams \& Lewis (1996) postulated that it was lensed gravitationally by a factor of $\mu$ $\sim$ 30, implying a SFR $<$ 100 \sfr. Pettini et al. (2000) took advantage of the high apparent brightness of cB58, and obtained deep rest-frame ultraviolet (UV) spectroscopy of cB58, which enabled the first quantitative study of the physical properties in high-redshift star-forming galaxies.  Additionally, Teplitz et al.\ (2000) obtained NIR spectroscopy, detecting numerous diagnostic emission lines.  Their measurements of the rest-frame optical emission lines implied a dust and lensing-corrected SFR of $\sim$ 20 \sfr, and Z $\sim$ $\frac{1}{3}$ Z\sol.

Similarly, Rowan-Robinson (1991) discovered the galaxy FSC 10214+4724, at z=2.286, originally thought to be intrinsically the brightest known galaxy.  This galaxy was later shown to be lensed by a factor of $\mu$ $\sim$ 100 (Eisenhardt et al.\ 1996).  Even with this strong lensing, this object still has an intrinsic luminosity consistent with an ultra-luminous infrared galaxy (ULIRG; L $>$ 10$^{12}$ L\sol).  The bright infrared luminosity implies the presence of a large amount of dust.  Teplitz et al.\ (2006) obtained {\it Spitzer Space Telescope} Infrared Spectrograph (IRS) spectroscopy in order to investigate the relative contribution between star formation and active galactic nucleus (AGN) activity.  They concluded that while an obscured AGN was present, star formation dominates the bolometric luminosity, although typical starburst polycyclic aromatic hydrocarbon (PAH) emission is strangely absent.  This object highlights the usefulness of strongly lensed galaxies, as does recent work by Siana et al.\ (2008, 2009), who have used IRS (as well as MIPS) to study the mid- and far-infrared properties of cB58 and the Cosmic Eye (z=3.074; $\mu$ = 28).

Recently, using the Sloan Digital Sky Survey (SDSS; York et al.\ 2000) Data Release 4 (DR4; Adelman-McCarthy et al.\ 2006), Allam et al.\ (2007) serendipitously discovered a galaxy\footnote[1]{While the 8 o'clock arc was not selected as a LBG using the photometric criteria of Steidel et al. (1996), the SDSS u$^{\prime}$, g$^{\prime}$, and r$^{\prime}$ fluxes of each component are consistent with a LBG, not accounting for bandpass differences.} with a brighter apparent magnitude, which they dubbed the 8 o'clock arc (denoting its time of discovery).  The total magnitude of all of the components of the 8 o'clock arc is i$^{\prime}$ = 19.0 mag, $\approx$ 1.4 mag brighter than cB58.  At z = 2.73 (a redshift nearly equal to that of cB58), the 8 o'clock arc is lensed by a luminous red galaxy at z = 0.38, resulting in a total magnification factor of $\mu$ = 12.3$^{+15}_{-3.6}$ (Allam et al. 2007).  The lensing distorts the galaxy into three separate components which form a partial ring, extending an angle of 162$^{o}$ and a total length of 9.6\arcs.  Allam et al.\ dub these components from west to east: A1, A2 and A3 (with i$^{\prime}$ = 20.13, 20.11 and 20.21, respectively), each of which are magnified by $\mu$ $\sim$ 4.  Allam et al.\ obtained deeper optical (rest-frame UV) imaging and spectroscopy, which confirmed the lensing nature of the object, and showed that its spectrum was comparable to LBGs at this redshift.  The redshift of 2.73 was computed from a number of absorption lines, including Ly$\alpha$ absorption.  While cB58 is a typical L$^{*}$ LBG, the 8 o'clock arc is intrinsically much brighter, with L $\sim$ 11L$^{*}$.

We present the results of a study on the physical properties of this intrinsically very bright galaxy, using optical spectroscopy to investigate the redshift, and NIR spectroscopy to measure the SFR and ionization properties.  In \S 2 we detail our spectroscopic observations, data reduction and emission line measurements.  In \S 3 we present our results and discussion, and in \S 4 we give our conclusions.  Where applicable, we use a cosmology with H$_{o}$ = 70 km s$^{-1}$ Mpc$^{-1}$, $\Omega_{m}$ = 0.3 and $\Omega_{\lambda}$ = 0.7.
\begin{figure*}[th]
\epsscale{1.15}
\plottwo{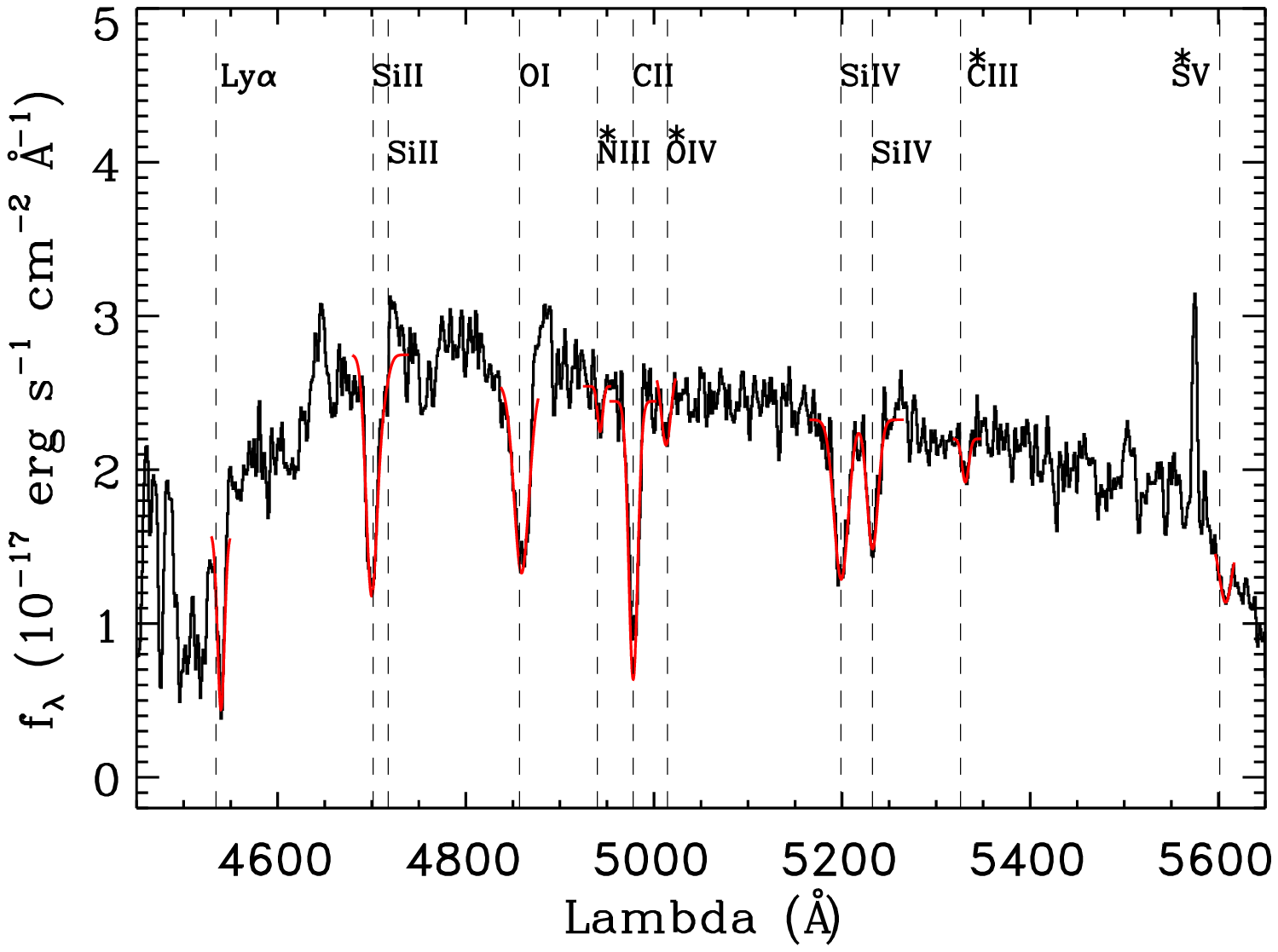}{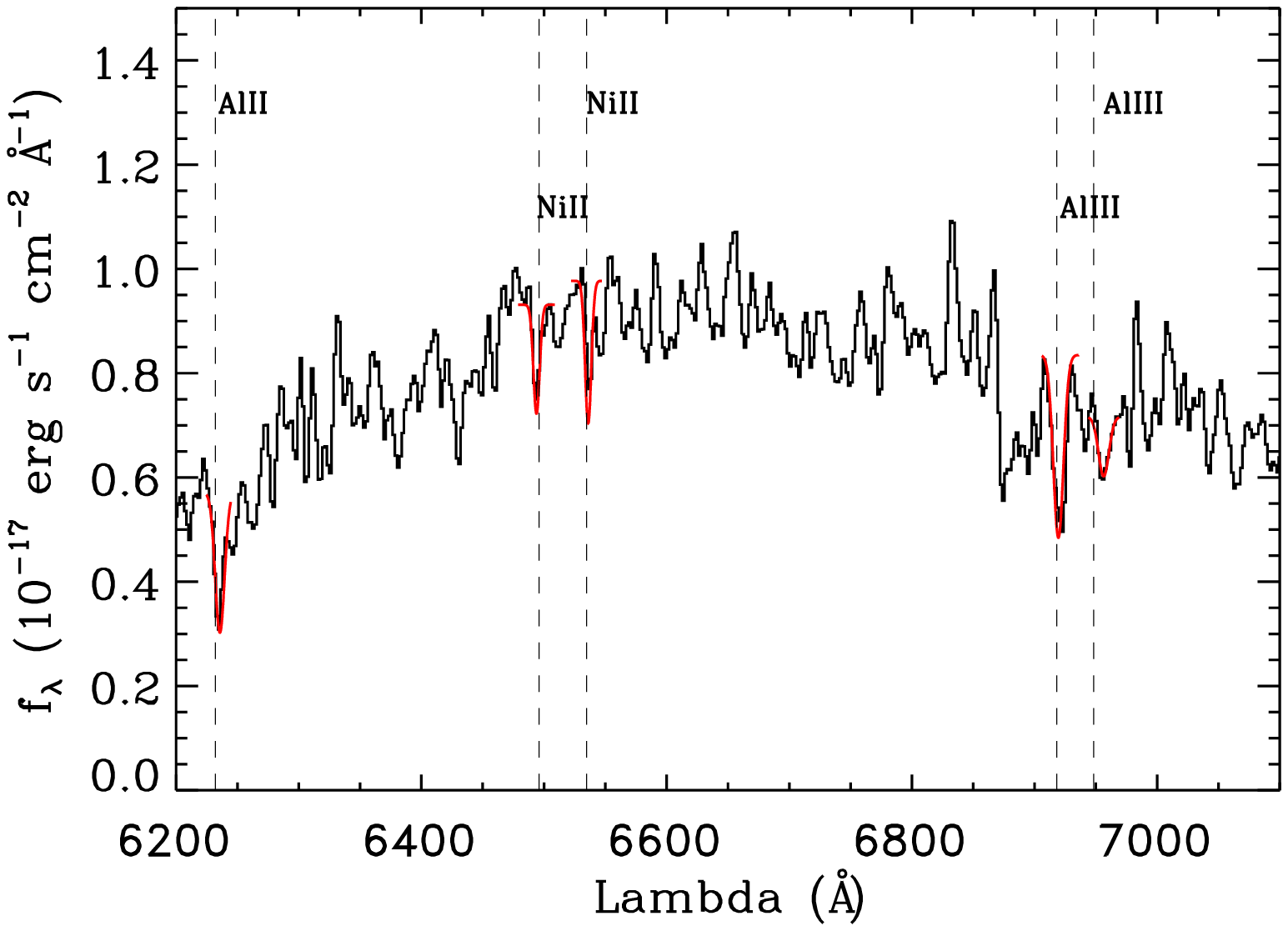}
\caption{Portions of our LRIS optical (rest-UV) blue and red-side spectra of the 8 o'clock arc, highlighting most of the lines which we measured.  Dotted lines show where the listed absorption features should fall assuming z = 2.73.  The red curves show the best-fit Gaussians to the observed features.  The asterisks denote the stellar photospheric absorption lines.}
\end{figure*}

\section{Observations}

\subsection{Optical Spectroscopy}
We obtained spectroscopy of the 8 o'clock arc with the low-resolution imaging spectrograph (LRIS; Oke et al.\ 1995; Steidel et al.\ 2004), combined with the LRIS atmospheric dispersion compensator (ADC) at the W.\ M.\ Keck Observatory on UT 2008 Jan 6.  We obtained 1200 s of data on the LRIS blue side with the 400/3400 grism, covering $\lambda\approx 3200-5600$~\AA.  We simultaneously obtained 1200 s of data on the LRIS red side with the 400/8500 grating covering $\lambda\approx 5600-8000$~\AA.  The characteristic seeing during the observations was 0.8\arcs.  We used a 1$^{\prime\prime}$ slit , oriented to acquire optimally the A1 and A3 components of the 8 o'clock arc.  The slit also achieves partial coverage of component A2.  We reduced the spectra using standard and custom IRAF\footnote[2]{IRAF is distributed by the National Optical Astronomy Observatory (NOAO), which is operated by the Association of Universities for Research in Astronomy, Inc.\ (AURA) under cooperative agreement with the National Science Foundation.} tasks (including custom tasks from D.~Stern 2006, private communication).   

The wavelength solution was found using lines from argon, cadmium, mercury, neon and zinc lamps (cadmium and zinc were included for their blue lines).  On the blue side, 16 lines were fit over 3300 -- 5460 \AA, with a residual rms of 0.3 \AA.  On the red side, 45 lines (primarily neon and argon) were fit over the whole wavelength range, with an rms of 0.42 \AA.  On both sides, the wavelength solution residuals were roughly constant as a function of wavelength, i.e. the solution did not degrade near the ends.  The spectra were calibrated using observations of the standard stars HZ~15 and PG~0310+149 taken on the night of observation.  However, the spectra of the 8 o'clock arc were obtained at airmass $>$2 while the standards have airmass 1.3--1.4, and some residual flux calibration issues are apparent near the edges of the wavelength coverage for both the blue and red sides.  Figure~1 shows a portion of both the blue and red-side LRIS spectra.   These spectra includes many strong ISM absorption lines as well as stellar coronal lines.

\subsection{Near-Infrared Spectroscopy}
\subsubsection{Data Acquisition and Reduction}
We obtained NIR H and K-band long-slit spectroscopy of the 8 o'clock arc with the Near InfraRed Imager and Spectrometer (NIRI; Hodapp et al.\ 2003) on the Gemini North 8m telescope.  We used a slit width of 0.72\arcs, and the slit was oriented to cover both the A2 (RA 00$^{h}$22$^{m}$40.97$^{s}$, $\delta$14$^{o}$31$^{\prime}$14.0\arcs~[J2000]) and A3 components (RA 00$^{h}$22$^{m}$41.15$^{s}$, $\delta$14$^{o}$31$^{\prime}$12.6\arcs~[J2000]; see Figure 2).  These data were taken as part of the Gemini queue over multiple nights, with an ABA$^{\prime}$B$^{\prime}$ dither pattern, in order to obtain the best possible conditions for this field and our science requirements.  The H-band spectra consist of 48 $\times$ 300 second exposures taken over three nights in July and August 2008, for a total exposure time of 4.0 hours.  The K-band spectra consist of 34 $\times$ 300 second exposures taken over five nights in June, July and August 2008, for a total exposure time of 2.8 hours.  Figure 2 displays a NIRI H-band image of the 8 o'clock arc, which was made using the 11 $\times$ 15 second H-band acquisition images, showing the locations of our slit, as well as labeling the three components of the arc.  Measurements of the sky emission lines in the raw frames give a spectral resolution of $\approx$ 33 \AA~in the H-band, and $\approx$ 43 \AA~in the K-band, for R $\sim$ 500.

These data were reduced using a combination of standard IRAF tasks, as well as custom NIRI tasks available in the Gemini external package to IRAF.  All raw data were first run through the Gemini task {\tt nprepare}, which updates the header keywords.  We then made a master flat field image for each night using {\tt nsflat}, which combines and normalizes the individual flat exposures.  The science images for each night (both the 8 o'clock arc and a standard star) were then divided by this image to remove flat field effects.  Each frame contained numerous cosmic ray hits, which we flagged with the spectroscopic version of the LACOSMIC routine (van Dokkum 2001).  The sky was removed from each image by subtracting a temporally adjacent exposure.  In some of the sky-subtracted images, we noticed strong vertical striping.  This is a known issue with NIRI, and the Gemini package has a task, {\tt nvnoise}, to correct this.  However, this task did not adequately remove all of these features in our data, so we excluded problematic images from the final image stack (30 min lost in the K-band, none in the H-band).  In addition, NIRI has a dark current instability when changing exposure types, thus we also excluded the first frame in each new sequence from the final stack.  None of these bad frames were used as sky frames.  Our total exposure times of 4.0 and 2.8 hours for the H and K-bands, respectively, do not include these bad frames.

We used the Gemini tasks {\tt nswavelength} and {\tt nstransform} to derive a wavelength calibration and rectify the data using argon lamp spectra.  The typical rms of the wavelength fit was $\sim$ 0.4 \AA~in the H-band, and 0.8 \AA~in the K-band.  Final two dimensional spectra for each night were made using the task {\tt nscombine}, using bad pixel masks created from a combination of the flat field and the cosmic ray rejection results.  Residual sky lines were removed using the task {\tt nsressky}.  In some cases (in the H-band), LACOSMIC flagged a large region in the upper corner.  While this did not affect the image near the 8 o'clock arc, it did affect the results of {\tt nsressky}.  In these few instances, we removed the top portions of the two-dimensional (2D) spectrum with {\tt nscut} prior to running {\tt nsressky}, allowing it to run smoothly.
\begin{figure}
\epsscale{1.15}
\plotone{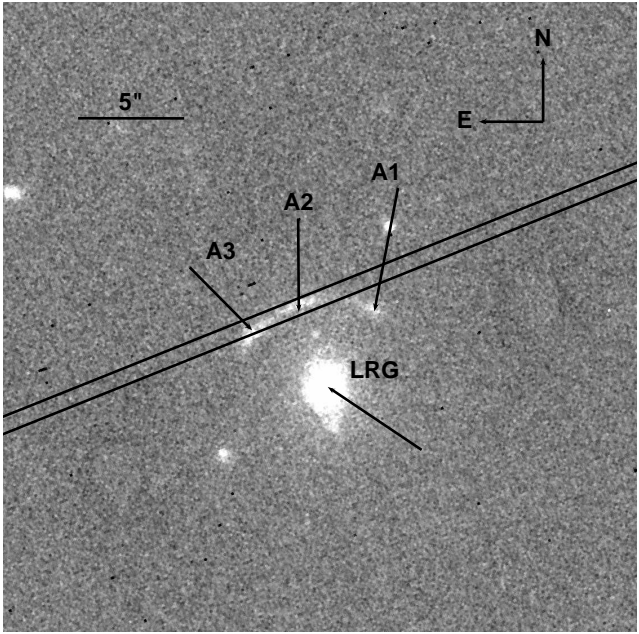}
\caption{NIRI H-band image of the 8 o'clock arc, made from combining the acquisition images.  The three components as defined by Allam et al.\ (2007) are labeled, and we show the position, orientation and width (0.72\arcs) of our slit, which covered components A2 and A3.  Also indicated is the luminous red galaxy (LRG) at z = 0.38 responsible for the lensing of the 8 o'clock arc.}
\end{figure}

\subsubsection{Spectral Extraction, Flux Calibration and Error Estimates}
One-dimensional spectra were traced and extracted from the 2D image for each night using the IRAF task {\tt apall}.  The trace was defined by the continuum, summing over 10 pixels at a time.  Visual inspection of the 2D spectra clearly revealed the two separate components.  We extract the sum of the two components to maximize the signal-to-noise, and later we will address differences between the two components by extracting them separately.  The size of the extraction aperture was based on the apparent spatial extent of the spectrum in the 2D image, which we measured to be 5.8\arcs~(50 pixels) in size (this was defined using H$\beta$ in the H-band and H$\alpha$ in the K-band, both of which are easily seen by eye in the 2D spectra; Figures 3 and 4).  The final 1D spectrum is a background-subtracted sum over the aperture.  Telluric absorption was corrected using the Gemini task {\tt nstelluric}, which was run interactively to ensure that the strong spectral features did not adversely affect the continuum fitting.  We used the standard HIP115986, of spectral type F0V.  We chose this telluric standard as it is a compromise between the strong hydrogen lines present in hotter stars, and other atomic lines seen in cooler stars.  Flux calibration was done by scaling a Kurucz 1993 F0V star model up to have a magnitude equal to the published 2 Micron All Sky Survey (2MASS) magnitude, which we converted to AB magnitudes of 11.483 and 11.876 in the H and K-bands, respectively.  This calibrated model was then divided by the spectrum of our (telluric-corrected) standard to create a calibration array, which was then multiplied into the object's spectrum.

We noticed that out of the three nights when H-band spectroscopy was obtained, one night (2008 Aug 21) had much higher quality data than the other two nights (2008 Jul 4 and 2008 Jul 5).  Examining the 2D spectra from 2008 Aug 21 by eye, we could clearly see all expected emission lines, while they remained much harder to pick out in the other two nights.  In addition, when examining the H$\beta$ emission line, we found a total signal-to-noise (S/N) = 26 on 2008 Aug 21, while the S/N = 5 and 4 on 2008 July 4 and 2008 Jul 5, respectively, which is significantly different than what we would expect from exposure time differences.  Examining the telluric standards for each of the three nights, the two July nights showed a much higher atmospheric opacity at the red end of the spectrum.  During the telluric correction step, these already weakly observed lines would then become even more uncertain as they were boosted to account for greater telluric absorption.  Rather than introduce these high multiplicative uncertainties into our final stack, we decided to move forward with only the data from 2008 Aug 21, which has a total exposure time of $\sim$ 1.6 hours.
\begin{deluxetable}{ccccc}
\tabletypesize{\small}
\tablecaption{Measured Absorption lines from LRIS Spectroscopy of the Eight O'clock Arc}
\tablewidth{0pt}
\tablehead{
\colhead{Absorption Line} & \colhead{Rest $\lambda$} & \colhead{Observed $\lambda$} & \colhead{Redshift} & \colhead{Line Type}\\
\colhead{$ $} & \colhead{(\AA)} & \colhead{(\AA)} & \colhead{$ $} & \colhead{$ $}\\
}
\startdata
Ly$\beta$\phantom{\ddag}&1025.18&3822.53&2.7286 $\pm$ 0.0002&ISM\\
Ly$\alpha$\phantom{\ddag}&1215.67&4539.66&2.7343 $\pm$ 0.0002&ISM\\
Si\,{\sc ii}\phantom{\ddag}&1260.42&4699.62&2.7286 $\pm$ 0.0004&ISM\\
Si\,{\sc ii}$^{\dag}$\phantom{\dag}&1264.74&4711.50&2.725\phantom{0} $\pm$ 0.001\phantom{0}&ISM\\
O\,{\sc i}\phantom{\ddag}&1302.17&4859.20&2.7316 $\pm$ 0.0005&ISM\\
N\,{\sc iii}\phantom{\ddag}&1324.32&4942.74&2.732\phantom{0} $\pm$ 0.002\phantom{0}&Stellar\\
C\,{\sc ii}\phantom{\ddag}&1334.53&4977.92&2.7301 $\pm$ 0.0001&ISM\\
O\,{\sc iv}\phantom{\ddag}&1344.30&5012.82&2.729\phantom{0} $\pm$ 0.001\phantom{0}&Stellar\\
Si\,{\sc iv}\phantom{\ddag}&1393.76&5199.04&2.7302 $\pm$ 0.0005&ISM\\
Si\,{\sc iv}\phantom{\ddag}&1402.77&5232.40&2.7300 $\pm$ 0.0006&ISM\\
C\,{\sc iii}\phantom{\ddag}&1427.85&5331.06&2.734\phantom{0} $\pm$ 0.002\phantom{0}&Stellar\\
S\,{\sc v}\phantom{\ddag}&1501.76&5607.77&2.7341 $\pm$ 0.0009&Stellar\\
C\,{\sc iv}$^{\ddag}$&1549.49&5779.40&2.7299 $\pm$ 0.0005&ISM\\
Al\,{\sc ii}\phantom{\ddag}&1670.79&6235.70&2.7322 $\pm$ 0.0002&ISM\\
Ni\,{\sc ii}\phantom{\ddag}&1741.55&6493.88&2.7288 $\pm$ 0.0002&ISM\\
Ni\,{\sc ii}\phantom{\ddag}&1751.92&6536.03&2.7308 $\pm$ 0.0001&ISM\\
Al\,{\sc iii}\phantom{\ddag}&1854.72&6919.46&2.7307 $\pm$ 0.0001&ISM\\
Al\,{\sc iii}\phantom{\ddag}&1862.79&6956.38&2.7344 $\pm$ 0.0004&ISM\\
\enddata
\tablecomments{$^{\dag}$ The Si\,{\sc ii} lines are extremely close together, thus we forced the FWHM of the weaker, redder line to match that of the bluer line.  $^{\ddag}$ The C\,{\sc iv} doublet was blended, thus we took the redshift of the line to be the fitted wavelength divided by the mean of the two rest wavelengths.  We computed the mean redshift by taking the weighted mean of all of the lines, where the error on the mean is the square root of inverse of the sum of the inverse variances.  The weighted mean redshift of all lines is 2.7308 $\pm$ 0.0006. Considering only stellar photospheric lines, we find z = 2.7322 $\pm$ 0.0012, while for ISM lines we find z = 2.7302 $\pm$ 0.0006.}

\end{deluxetable}

To estimate the errors in our spectra, we first made a 2D noise image for each night, where each pixel had a value equal to the standard deviation of the mean of values in that pixel from all input images from that night.  We then made a 1D error spectrum using the same aperture width as for the science data, but adding rows in quadrature.  These 1D error spectra were then propagated through the subsequent steps of telluric correction and flux calibration.  We were thus left with a 1D object spectrum and a 1D noise spectrum for each night.

The 1D spectra from each night for were averaged together (for each band), using the statistical weighted mean, to create a final spectrum.  Similarly, a final noise spectrum was created by taking the square root of the variance of the weighted mean.  In Figures 3 and 4 we show the final 1D spectra, along with the 2D spectra and the sky noise from one night in each band, respectively.

As discussed in \S 3.5, we also reduced the acquisition images, which consisted of data in the NIRI H and K$^{\prime}$ filters.  These data were reduced by combining standard IRAF tasks with tasks in the gemini package.  The final reduced images were flux calibrated using 2MASS stars in the field.

\section{Results}

\subsection{Redshift}
\subsubsection{Absorption Line Measurements}
Pettini et al. (2000 \& 2002) published tables of stellar photospheric and ISM absorption lines detected in the rest-frame UV of cB58.  Given the similarity in redshift, we might expect to detect many of these lines in our LRIS spectra, although our reduced exposure time will result in some non-detections.  We detect 18 absorption lines, of which four are stellar photospheric in nature, and 14 come from the ISM.  The lines range from hydrogen to nickel, with ionization states from H\,{\sc i} to S\,{\sc v}.
\begin{figure}[th]
\epsscale{1.15}
\plotone{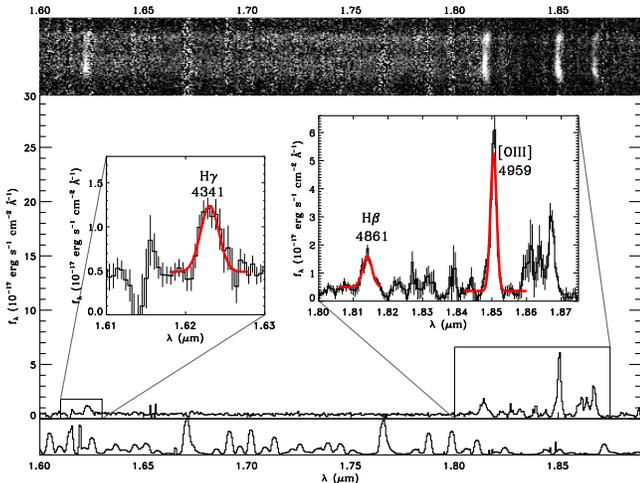}
\vspace{1mm}
\caption{Gemini/NIRI H-band spectrum of the 8 o'clock arc is shown in the main panel.  The two insets are drawn to highlight our regions of interest.  In the left inset, we show the H$\gamma$ emission line, with the Gaussian fit shown in red.  Likewise, in the right inset we show fits to H$\beta$ and [O\,{\sc iii}] $\lambda$4959.  The top panel shows the 2D spectrum, which shows that each of the emission lines are detectable by eye (including [O\,{\sc iii}] $\lambda$5007; see \S 3.3).  The top component corresponds to A3, while the more extended, bottom component corresponds to A2.  The lower panel shows the sky emission as measured from one of the raw spectroscopic frames.  The decrease in transmission redward of 1.8 $\mu$m is due to increasing atmospheric opacity.}
\end{figure}

We used IDL with the MPFIT package and the MPFITFUN task to fit Gaussian absorption lines to each of these 19 line positions.  During the fitting, the line widths were limited to be greater than (or equal to) the resolution of the data, which was 6.5 and 6 \AA~for the blue and red sides, respectively.  Table 1 gives the results of these fits, with the errors on the individual redshifts coming from 10$^{3}$ Monte Carlo simulations, where we varied the flux in each wavelength bin by an amount proportional to the flux errors.  Figure 1 shows a portion of our blue and red channel spectra, showing the positions and Gaussians for the lines which we fit.  The mean redshift was computed by taking the weighted mean of the lines, where the weights are the inverse variances of each line.  The corresponding error on the mean is the square root of inverse of the sum of the weights.  The mean redshift of all of the absorption lines is z = 2.7308 $\pm$ 0.0006.  

The mean redshift is a combination of the stellar photospheric lines z$_{*}$, which we expect to reside at the systemic redshift of the galaxy, and the ISM lines z$_{ISM}$, which may be at a slightly different redshift depending on the kinematics of the galaxy.  Separating these out, we find z$_{*}$ = 2.7322 $\pm$ 0.0012 and z$_{ISM}$ = 2.7302 $\pm$ 0.0006.  This blueshift of the ISM lines implies that we are observing an outflow with v $\sim$ 160 km s$^{-1}$.  This may not be surprising given the high SFR we calculate in \S 3.2, as strong outflows on the order of hundreds of km s$^{-1}$ have been observed in star-forming galaxies locally (Kunth et al.\ 1998), as well as at high redshift (e.g., Shapley et al.\ 2003).  It is of note that the ISM redshift measurement does not include Ly$\alpha$ and Ly$\beta$, as these lines appeared at significantly different redshifts.

\subsubsection{Emission Line  Measurements}
The MPFITFUN task was also used to obtain Gaussian line flux measurements from the NIRI spectra.  We only fit lines which we could detect by eye, which were H$\gamma$ ($\lambda$4341), H$\beta$ ($\lambda$4861) and [O\,{\sc iii}] ($\lambda$4959) in the H-band, and H$\alpha$ ($\lambda$6563) in the K-band.  While [O\,{\sc iii}] ($\lambda$5007) is clearly detected by eye in the 2D spectrum, it lies in a region of low atmospheric transmission.  Its telluric correction is large and any measurements from this line are uncertain, thus we exclude it from further analysis.
\begin{figure}[th]
\epsscale{1.15}
\plotone{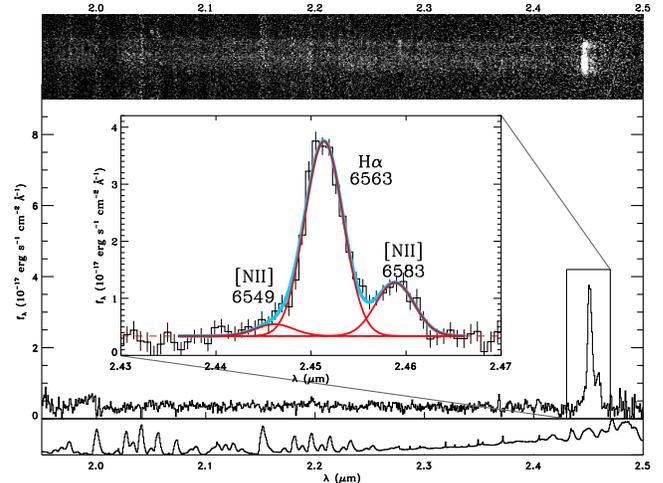}
\vspace{1mm}
\caption{Same as Figure 3, only for the K-band.  The red curves show the individual Gaussian fits to each of the three emission lines, while the blue curve draws the combination of the three.  Although the thermal infrared becomes more of a problem at $\sim$ 2.45 $\mu$m, we are still able to detect H$\alpha$ at very high S/N.}
\end{figure}

 While the H-band lines are separated enough in wavelength that single Gaussian fits are sufficient, at our resolution H$\alpha$ is blended in with two [N\,{\sc ii}] lines, $\lambda$6549 and $\lambda$6583.  We thus fit these lines with a triple Gaussian, fitting iteratively to fix the redshifts and widths of the [N\,{\sc ii}] Gaussians to match that of H$\alpha$ (z = 2.7351 \AA~and observed FWHM = 47.6 \AA, respectively).  While MPFITFUN does compute an error on the fit, it is more indicative of how well a Gaussian fits the spectrum rather than the integrated photometric errors along the curve.  We thus computed a line flux error via running 10$^{3}$ Monte Carlo simulations, again allowing each data point (both the continuum and the line) to vary by an amount proportional to its flux error (we note that the uncertainties on H$\alpha$ and [N\,{\sc ii}] will be covariant).  Table 2 gives the fitting results for each line we measured, accounting for the effects of galactic extinction (Schlegel et al.\ 1998).  We assume that stellar absorption in the Balmer lines is negligible for H$\alpha$, and that the effect is of equal magnitude for all three Balmer lines (Olofsson 1995).  From Table 2, the weighted mean redshift of our measured emission lines is z$_{em}$ = 2.7333 $\pm$ 0.0001, which is consistent with the systemic redshift derived from the stellar photospheric absorption lines.
\begin{deluxetable*}{cccccc}
\tabletypesize{\small}
\tablecaption{Emission Line Results from NIRI Observations of the Eight O'clock Arc}
\tablewidth{0pt}
\tablehead{
\colhead{Extraction} & \colhead{Line} & \colhead{Observed $\lambda$} & \colhead{Line Flux} & \colhead{FWHM} & \colhead{Redshift}\\
\colhead{$ $} & \colhead{Rest $\lambda$ (\AA)} & \colhead{(\AA)} & \colhead{(10$^{-17}$ erg s$^{-1}$ cm$^{-2}$)} & \colhead{(\AA)} & \colhead{$ $}\\
}
\startdata
Total&H$\gamma~\lambda$4341&16230.5&22.44 $\pm$ 4.79&27.3&2.739\\
CompA2&$ $&16229.0&15.73 $\pm$ 3.50&24.6&2.739\\
CompA3&$ $&16235.6&7.69 $\pm$ 8.03&30.1&2.740\\
Total&H$\beta~\lambda$4861&18140.5&37.38 $\pm$ 5.58&31.6&2.732\\
CompA2&$ $&18140.4&27.92 $\pm$ 3.90&30.0&2.732\\
CompA3&$ $&18146.6&15.95 $\pm$ 7.90&56.5&2.733\\
Total&$[$O\,{\sc iii}$]~\lambda$4959&18504.8&111.96 $\pm$ 5.39&21.0&2.732\\
CompA2&$ $&18504.1&78.31 $\pm$ 4.24&20.0&2.731\\
CompA3&$ $&18505.3&36.16 $\pm$ 3.84&18.6&2.732\\
Total&$[$N\,{\sc ii}$]~\lambda$6549&24461.3&10.74 $\pm$ 4.64&47.6&2.735\\
CompA2&$ $&24461.6&5.38 $\pm$ 3.93&48.3&2.735\\
CompA3&$ $&24469.1&4.02 $\pm$ 2.82&45.1&2.736\\
Total&H$\alpha~\lambda$6563&24513.6&176.06 $\pm$ 5.48&47.6&2.735\\
CompA2&$ $&24513.8&117.88 $\pm$ 4.22&48.3&2.735\\
CompA3&$ $&24521.4&48.25 $\pm$ 3.14&45.1&2.736\\
Total&$[$N\,{\sc ii}$]~\lambda$6583&24588.3&48.32 $\pm$ 3.92&47.6&2.735\\
CompA2&$ $&24588.6&31.88 $\pm$ 3.21&48.3&2.735\\
CompA3&$ $&24596.2&9.78 $\pm$ 2.37&45.1&2.736\\
\enddata
\tablecomments{Both [N\,{\sc ii}] lines were fixed to have the same redshift and FWHM as the H$\alpha$ line. While $[$O\,{\sc iii}$]~\lambda$5007 is detected by eye in the 2D spectrum, its position in a high atmospheric opacity region makes measurement of its flux unreliable, thus we only measure $[$O\,{\sc iii}$]~\lambda$4959. Errors were computed via Monte Carlo simulations. We sorted the simulations by their best-fit value, and consider the 1 $\sigma$ error as being the spread in best-fit values over the middle 68\%.  The weighted mean of the total galaxy's redshift is z$_{emission}$ = 2.7333 $\pm$ 0.0001.}

\end{deluxetable*}

\subsection{Dust Extinction and Star Formation Rate}
Equation 1 (Kennicutt 1998) details how to calculate the SFR from the H$\alpha$ luminosity, assuming a Salpeter initial mass function (IMF) over 0.1 -- 100 M\sol.

\begin{equation}
SFR~(M_{\odot}~yr^{-1}) = 7.9 \times 10^{-42}~\frac{L_{H\alpha}}{erg~s^{-1}}
\end{equation}

This luminosity can be computed from the line flux given the redshift of the galaxy.  Our measurement of f$_{H\alpha}$ = 1.76 $\pm$ 0.05 $\times$ 10$^{-15}$ erg s$^{-1}$ cm$^{-2}$ gives L$_{H\alpha}$ = 1.09 $\pm$ 0.03 $\times$ 10$^{44}$ erg s$^{-1}$, which corresponds to an observed SFR of 863 $\pm$ 27 \sfr~(uncorrected for the gravitational lensing magnification).  However, this is only a lower limit, as it neglects dust obscuration.  We have the data necessary to compute the Balmer decrement, which can allow us to infer the dust extinction by comparing Balmer line ratios to their predicted values via atomic physics.  The color excess can then be computed via Equation 2, 
\begin{equation}
E_{g}(B-V) = 2.5~\frac{log\left(f_{H\alpha}/f_{H\beta}/2.86\right)}{k^{\prime}(\lambda_{H\beta}) - k^{\prime}(\lambda_{H\alpha})}
\end{equation}
where E$_{g}$(B-V) is the gas phase color excess, 2.86 is the intrinsic $f_{H\alpha}/f_{H\beta}$ ratio (for Case B recombination, T$_{e}$ = 10$^{4}$ K and n$_{e}$ = 10$^{2}$; Osterbrock 1989), and k$^{\prime}$($\lambda$) is the Calzetti et al.\ (2000) starburst reddening curve value for nebular gas at wavelength $\lambda$.

Using Equation 2, we computed the gas phase color excess E$_{g}$(B-V)$_{\alpha/\beta}$ = 0.97 $\pm$ 0.30.  Due to the bright intrinsic nature of this galaxy as well as the lensing magnification, we also detect H$\gamma$ emission, which is very rare for this high redshift.  This gives another estimate of the color excess using the intrinsic $f_{H\alpha}/f_{H\gamma}$ ratio of 6.11, of E$_{g}$(B-V)$_{\alpha/\gamma}$ = 0.34 $\pm$ 0.30.  We also computed E$_{g}$(B-V)$_{\beta/\gamma}$ = 1.17 $\pm$ 1.22 from the ratio of $f_{H\beta}/f_{H\gamma}$, but the higher uncertainty in the fluxes of H$\beta$ and H$\gamma$ results in less confidence in this measurement.  The discrepancy between these measurements may not be surprising, as $f_{H\alpha}/f_{H\gamma}$ may miss more deeply embedded star formation, while H$\beta$ is observed near an atmospheric opacity trough.  

We thus define a final color excess measurement, a weighted mean of the three preceding measurements, of E$_{g}$(B-V) = 0.67 $\pm$ 0.21.  From observations of local galaxies, Calzetti et al.\ (2000) show that nebular emission lines typically suffer higher extinction that the stellar continuum, and that the color excess of the stellar continuum can be found via E$_{s}$(B-V) = (0.44 $\pm$ 0.03) E$_{g}$(B-V).  Extrapolating this to high redshift, we calculate E$_{s}$(B-V) = 0.29 $\pm$ 0.09, which corresponds to an extinction of A$_{5500}$ = 1.17 $\pm$ 0.36 mag.  

We can then correct the H$\alpha$ flux for dust extinction and derive an extinction-corrected SFR.  The intrinsic H$\alpha$ flux is found via f$^{\prime}_{H\alpha}$ = f$_{H\alpha}$10$^{0.4 E_{g}(B-V)k^{\prime}(\lambda_{H\alpha})}$, which yields f$^{\prime}_{H\alpha}$ = 4.34 $\pm$ 1.22 $\times$ 10$^{-15}$ erg s$^{-1}$ cm$^{-2}$, more than two times the uncorrected value.  The extinction-corrected SFR is thus 2129 $\pm$ 595 \sfr~(where $\sim$ 90\% of this error comes from the uncertainty in the dust extinction).  However, this does not include a correction for lensing.  Allam et al.\ (2007) found the total lensing magnitude to be $\mu$ = 12.3, where each of the three individual components had $\mu$ $\sim$ 4, assuming each component is a counter image of the intrinsic source.  This implies that each component represents a lensed image of the whole galaxy.  

However, we find significant differences in the properties (i.e. SFR, stellar mass, E[B-V]) of the different components, and we conclude that the lensing geometry is more complicated than this simple configuration.  Nevertheless, we assume that each component is lensed by $\mu$ = 4, adopting the Allam et al.\ results.  We note that if this is not the case, only our delensed SFR and our mass estimates (\S 3.5) are affected.  Given that our total spectrum combines two components, we need to correct for a lensing factor of $\sim$ 8; assuming this value, we find a dust and lensing-corrected SFR = 266 $\pm$ 74 \sfr.  Allam et al.\ give an estimated SFR $\sim$ 230 \sfr~(accounting for our cosmology) as a rough estimate for the 8 o'clock arc, based on scaling relations to cB58.  Although there are significant differences in extinction and physical properties between the 8 o'clock arc and cB58, the SFR we derive is consistent.  These results are summarized in Tables 3 and 4.
\begin{figure}
\epsscale{1.2}
\plotone{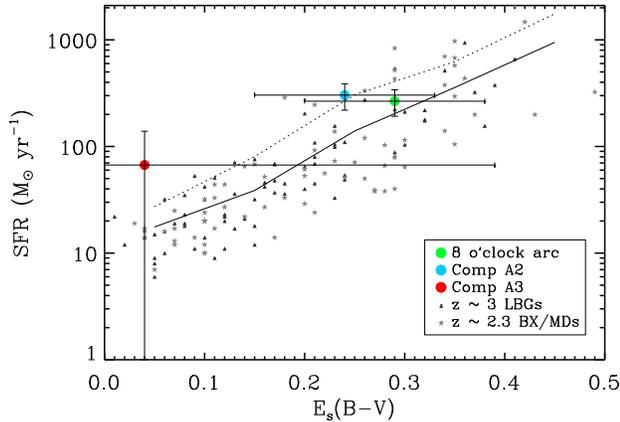}
\caption{We plot the SFR and stellar continuum color excess measured for the total 8 o'clock arc and its two separate components versus two samples of high-redshift galaxies.  Shown as triangles, the z $\sim$ 3 LBGs come from Shapley et al.\ (2001), while the z $\sim$ 2.3 BX/MD selected galaxies, shown as stars, come from Shapley et al.\ (2005a).  Both of these studies derived the SFRs and color excesses via stellar population modeling.  The solid and dotted line represent the mean and 1 $\sigma$ variation of the combined Shapley et al.\ galaxies, respectively.}
\end{figure}

In order to understand the significance of the extinction corrected star-formation rate, we compared our results to two samples of high-redshift star-forming galaxies.  First, we used the sample of z $\sim$ 3 LBGs from Shapley et al.\ (2001), where they derived both color excess and SFR values via stellar population model fitting of a subsample of spectroscopically confirmed LBGs, using both rest-frame UV and optical data.  They found a range of E(B-V) from 0 to 0.4, and SFRs from 5 up to $\sim$ 950 \sfr, with a clear trend of increasing SFR with higher values of E(B-V).  Comparing our results from the 8 o'clock arc to this sample, we find that it is in the 85th percentile of color excess, and 91st percentile of SFRs.  Additionally, when compared to a sample of z $\sim$ 2.3 BX/MD selected galaxies in Shapley et al.\ (2005a), we find that the 8 o'clock arc is in the 78th percentile of color excess, and 79th percentile of SFRs.  Figure 5 shows the 8 o'clock arc compared to these samples.  We can see that this object follows the observed trend in z $\sim$ 2--3 galaxies, and it is among the top 20\% of star-forming galaxies at these redshifts.

\subsection{Metallicity}
Emission lines are commonly used to estimate gas-phase metallicities in extragalactic HII regions.  One of the most accurate estimates requires a measurement of the electron temperature, which one can calculate by the ratio of the auroral [O\,{\sc iii}] $\lambda$4363 to the nebular [O\,{\sc iii}] $\lambda\lambda$ 4959,5007 (Osterbrock 1989).  However, auroral lines are intrinsically faint and are undetected in high-redshift galaxies, even in other gravitationally lensed examples (e.g., Teplitz et al 2000).  Other metallicity measurements are available whereby one does not need the electron temperature, including two indices which we can measure with our data, N2 and O3N2.  The N2 index uses the [N\,{\sc ii}] $\lambda$6583/H$\alpha$ ratio to estimate the oxygen abundance (see Storchi-Bergmann et al.\ 1994).  However, as the N2 index saturates at Z\sol, Pettini \& Pagel (2004) have proposed the O3N2 index, which combines N2 with measurements of [O\,{\sc iii}] $\lambda$5007 and H$\beta$, providing a more accurate index for super-solar metallicities.  

Given our measurements, we can compute both the N2 and O3N2 indices.  We use the relations from Pettini \& Pagel (2004), 
\begin{equation}
12 + \log\left(O/H\right) = 8.90 + 0.57 \times N2
\end{equation}
\begin{equation}
12 + \log\left(O/H\right) = 8.73 - 0.32 \times O3N2
\end{equation}
where N2 is defined as log([N\,{\sc ii}] $\lambda$6583 / H$\alpha$) and O3N2 is log[([O\,{\sc iii}] $\lambda$5007 / H$\beta$)/([N\,{\sc ii}] $\lambda$6583 / H$\alpha$)].  In our spectrum of the 8 o'clock arc, [O\,{\sc iii}] $\lambda$5007 is plagued by strong telluric absorption.  We thus use [O\,{\sc iii}] $\lambda$4959 and estimate [O\,{\sc iii}] $\lambda$5007 using the predicted intrinsic ratio of f$_{5007}$ = 2.98 $\times$ f$_{4959}$ (Sorey \& Zeippen 2000).  All lines were corrected for dust extinction, as found in \S 3.2.  However, given that each of these measures uses ratios of lines which are close in wavelength, effects of the dust extinction on these ratios are small, and we neglect the uncertainty in dust extinction here.

The nebular emission line ratios can also probe for ionization by an AGN, by comparing the line ratios with, for example, a BPT diagram (Baldwin, Phillips \& Terlevich 1981), using [O\,{\sc iii}] $\lambda$5007/H$\beta$ versus [N\,{\sc ii}] $\lambda$6583/$H\alpha$, as shown in Figure 6.  In this figure, we plot the line ratios for the 8 o'clock arc over a sample of $\sim$ 10$^{5}$ star-forming galaxies and $\sim$ 3 $\times$ 10$^{3}$ AGNs from the SDSS, with both samples obtained from the CMU-PITT SDSS Value Added Catalog database\footnote[3]{http://nvogre.phyast.pitt.edu/vac/}.  We also show four z $\sim$ 2.3 galaxies from Erb et al.\ (2006a) and the theoretical maximum starburst line from Kewley et al.\ (2001).  From the [N\,{\sc ii}]/H$\alpha$ ratio alone, the 8 o'clock arc seems in line with the star-forming galaxies.  However, the [O\,{\sc iii}]/H$\beta$ ratio lies above the Kewley et al. maximum starburst line, although within 2 -- 3 $\sigma$ of the upper limit.  We nonetheless believe that our object is likely consistent with a star-forming galaxy, and that the discrepancy in the [O\,{\sc iii}] flux(es) could be due to uncertainty in our telluric correction, or possibly due to assumptions made by Kewley et al. (i.e. continuous star formation).  We cannot otherwise explain the factor of two higher [O\,{\sc iii}] flux compared with the Erb et al.\ objects.
\begin{figure*}
\epsscale{1.15}
\plotone{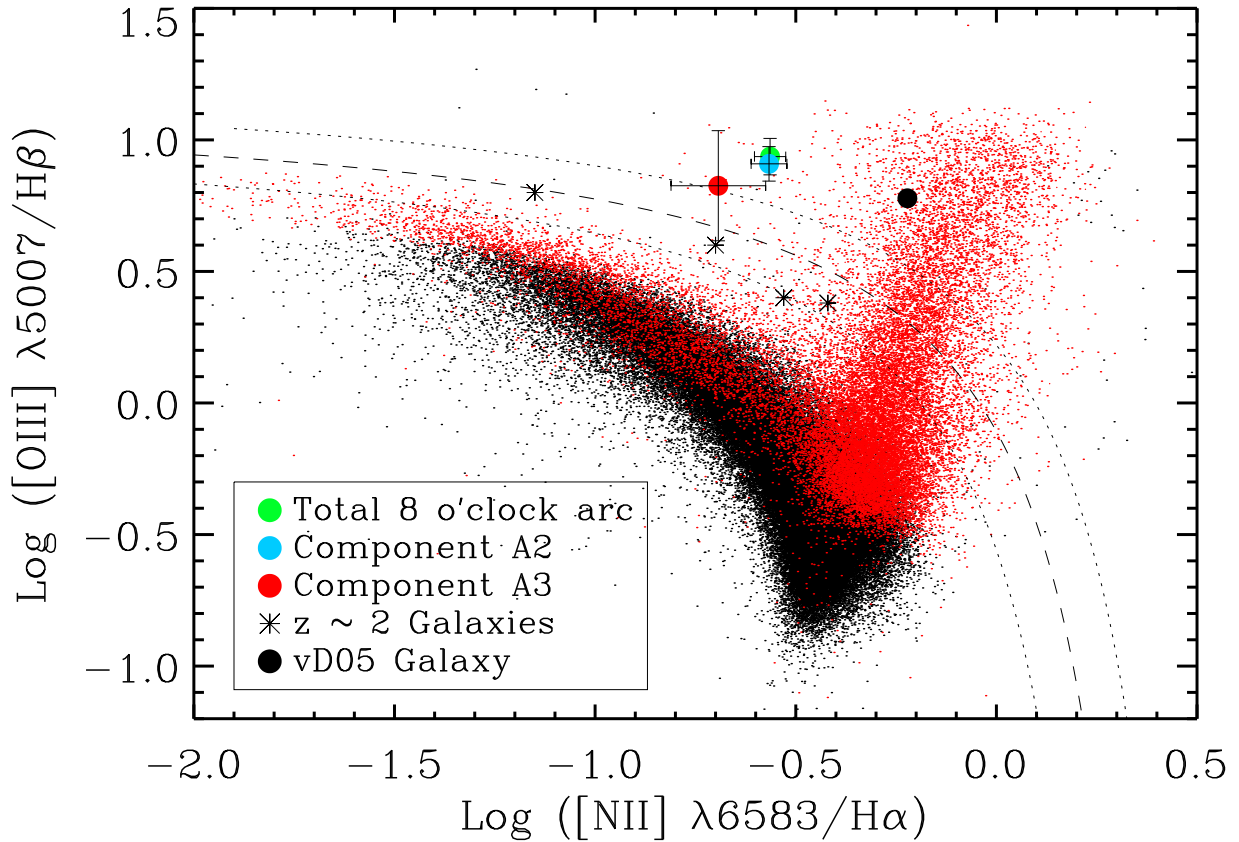}
\caption{The total 8 o'clock arc and both separate components are plotted on a line diagnostic ratio diagram in the manner of Baldwin, Phillips \& Terlevich (1981).  The [O\,{\sc iii}] $\lambda$5007 flux is scaled from the 4959 flux.  Small black dots denote $\sim$ 10$^{5}$ star-forming galaxies, while red dots denote $\sim$ 3 $\times$ 10$^{3}$ AGNs, both from the SDSS.  For comparison, we plot four z $\sim$ 2 star-forming galaxies from Erb et al.\ (2006a), as well as the shock-powered z = 2.225 galaxy from van Dokkum (2005).  The maximum starburst line from Kewley et al.\ (2001) is also shown as the dashed line, with the 1 $\sigma$ error on this line denoted by dotted lines.  While the 8 o'clock arc is 2 -- 3 $\sigma$ above the upper limit of this theoretical curve, this is likely due to uncertainties in the oxygen flux, or it could indicate that shocks contribute to the ionization.}
\end{figure*}

While we cannot rule out some AGN component to the [O\,{\sc iii}] line, the [N\,{\sc ii}]/H$\alpha$ ratio is consistent with expectations for star-forming galaxies with no apparent AGN contamination.  In addition, AGN typically have many rest-frame UV emission lines which should be present in our LRIS spectra, however no such lines are seen.  Additionally, the high [O\,{\sc iii}]/H$\beta$ ratio is seen in both components with no measurable spatial variation, while if an AGN was responsible for this high ratio, one would only expect to see it in the immediate vicinity of the galactic center.  Given these multiple lines of evidence, we find it highly unlikely that an AGN is dominating the emission from this galaxy.  However, the higher [O\,{\sc iii}]/H$\beta$ ratios could be indicative of photoionization due to strong galactic winds.  Van Dokkum et al. (2005) measured similar ratios in a luminous galaxy at z = 2.2 (also shown in Figure 6).  They found that their observed ratios could be be matched by a ``shock+precursor'' model (Dopita and Sutherland 1995), where the [N\,{\sc ii}]/$H\alpha$ ratio is produced by a shock, and the [O\,{\sc iii}]/H$\beta$ ratio is produced by photoionization of the preshock regions by UV and X-ray photons generated in the shock.  The shock itself is likely caused by a galaxy-wide wind from supernovae and the winds from massive stars, both of which imply ongoing star-formation.  Thus, even with a shock present, the observed emission lines come from some combination between nebular lines in H\,{\sc ii} regions, and photoionization from the shock.

Although the [N\,{\sc ii}]/$H\alpha$ ratio of the 8 o'clock arc is typical for star-forming galaxies, the high [O\,{\sc iii}]/H$\beta$ ratio necessitates a comparison with the shock models.  Figure 2 of van Dokkum et al.\ (2005) plot a BPT diagram similar to our own Figure 6.  They overplot shock only and shock+precursor models from Dopita and Sutherland (1995).  Placing the 8 o'clock arc on this diagram, we find that its line ratios put it at a lower [N\,{\sc ii}]/H$\alpha$ ratio than the shock+precursor models, although the [O\,{\sc iii}]/H$\beta$ ratio implies outflow velocities similar to what we measured in \S 3.1.1.  However, as we mentioned above, due to the sky emission and the increased telluric absorption around the observed wavelength of [O\,{\sc iii}], it is difficult to conclude that shock ionization is a potential ``contaminant'' of the emission lines.  We will therefore assume that the emission lines are dominated by star formation - if this is untrue, than our derived SFR is at worst an upper limit.

Using our dust corrected emission line fluxes, we have thus computed both the N2 and O3N2 indices\footnote[4]{For both the N2 and O3N2 indices, we include the intrinsic scatter in the relations in these error estimates, which is 0.18 and 0.14 dex, respectively.}.  First, from the N2 index, we find that 12 + log(O/H) = 8.58 $\pm$ 0.18, or (O/H) = 3.79 $\times$ 10$^{-4}$.  For a solar value of 12 + log(O/H) = 8.66 (Asplund et al.\ 2004), this corresponds to a gas-phase metallicity of Z = 0.83 $\pm$ 0.35 Z\sol.  

Using the O3N2 index, we find that 12 + log(O/H) = 8.25 $\pm$ 0.14, which corresponds to Z = 0.39 $\pm$ 0.13 Z\sol.  While this value is only $\sim$ half of the N2 metallicity, they are consistent within 1 $\sigma$.  Given the uncertainty in the [O\,{\sc iii}] flux, we regard the N2 derived metallicity as being more reliable, and thus the 8 o'clock arc has a gas-phase metallicity which is $\sim$ 80\% solar.

This gas-phase metallicity is consistent with other studies at 1 $\leq$ z $\leq$ 3.  Shapley et al.\ (2005b) studied the abundances with these same two estimators in DEEP2 star-forming galaxies at 1 $\leq$ z $\leq$ 1.4.  From the average of their 12 galaxies, they found Z $\sim$ 0.8 Z\sol~from the N2 index (Z $\sim$ 0.5 Z\sol~from the O3N2 index), with the two galaxies nearest in stellar mass to the 8 o'clock arc (see \S 3.5) having Z $\sim$ Z\sol.  With a sample of seven star-forming galaxies at 2.1 $<$ z $<$ 2.5, Shapley et al.\ (2004) found approximately solar metallicities using the N2 index, with stellar masses $\sim$ 10$^{11}$ M\sol~on average.  Using the N2 index to study the mass-metallicity relation at z $>$ 2, Erb et al.\ (2006a) found metallicities from $\sim$ 0.35 to 0.85 Z\sol~in a sample of z $\sim$ 2.3 star-forming galaxies.  Our results for the 8 o'clock arc place it near the upper end of these other high-redshift samples.  Given that the luminosity of the 8 o'clock arc is $\sim$ 11L$^{*}$, and a stellar mass of $\approx$ 2 $\times$ 10$^{11}$ M\sol~(see \S 3.5), we may expect it to have a metallicity higher than that for typical L$^{*}$ galaxies at this redshift (e.g., via the mass-metallicity relation; Erb et al. 2006a), an interpretation that is consistent with the 0.32 Z\sol~metallicity measured for the less-luminous L$^{*}$ galaxy cB58 (Teplitz et al. 2000).
\begin{deluxetable*}{cccccc}
\tabletypesize{\small}
\tablecaption{Measured Color Excesses}
\tablewidth{0pt}
\tablehead{
\colhead{Extraction} & \colhead{E$_{g}$(B-V)} & \colhead{E$_{g}$(B-V)} & \colhead{E$_{g}$(B-V)} & \colhead{E$_{g}$(B-V)} & \colhead{E$_{s}$(B-V)}\\
\colhead{$ $} & \colhead{$H\alpha/H\beta$} & \colhead{$H\alpha/H\gamma$} & \colhead{$H\beta/H\gamma$} & \colhead{Weighted Mean} & \colhead{Weighted Mean}\\
}
\startdata
Total&0.97 $\pm$ 0.30&0.34 $\pm$ 0.30&1.17 $\pm$ 1.22&0.67 $\pm$ 0.21&0.29 $\pm$ 0.09\\
Comp A2&0.76 $\pm$ 0.28&0.28 $\pm$ 0.31&0.87 $\pm$ 1.23&0.55 $\pm$ 0.20&0.24 $\pm$ 0.09\\
Comp A3&0.11 $\pm$ 0.97&0.04 $\pm$ 1.44&0.14 $\pm$ 5.42&0.09 $\pm$ 0.79&0.04 $\pm$ 0.35\\
\enddata

\end{deluxetable*}

\begin{deluxetable*}{ccccc}
\tabletypesize{\small}
\tablecaption{Measured Star Formation Rate and Metallicity}
\tablewidth{0pt}
\tablehead{
\colhead{Extraction} & \colhead{Dust Corr. SFR} & \colhead{De-Lensed SFR} & \colhead{Z$_{N2}$} & \colhead{Z$_{O3N2}$}\\
\colhead{$ $} & \colhead{M\sol~yr$^{-1}$} & \colhead{M\sol~yr$^{-1}$} & \colhead{(Z\sol)} & \colhead{(Z\sol)}\\
}
\startdata
Total&2129 $\pm$ 595&266 $\pm$ 74&0.83 $\pm$ 0.35&0.39 $\pm$ 0.13\\
Comp A2&1213 $\pm$ 337&303 $\pm$ 84& 0.82 $\pm$ 0.34&0.40 $\pm$ 0.13\\
Comp A3&266 $\pm$ 286&67 $\pm$ 72&0.70 $\pm$ 0.31&0.38 $\pm$ 0.14\\
\enddata

\end{deluxetable*}

\subsection{Physical Properties Across the Face of the Galaxy}
The lensing nature of the 8 o'clock arc gives us a unique opportunity to examine how these characteristics change across the face of the galaxy, as we oriented our slit such that it covered two components.  To extract these components separately, we used the same extraction procedure as before, only this time we split the extraction aperture into two parts, one covering the brighter component (A2; 3.6\arcs), and one covering the fainter component (A3; 2.2\arcs).  We then took these separate spectra, and fit emission lines to them as described in \S 3.1.2.  These results are also shown in Table 2.

We find significant differences in dust extinction in the different components.  From the Balmer line ratios, we measure the color excesses, with E$_{g}$(B-V) = 0.0.55 $\pm$ 0.20 (0.09 $\pm$ 0.79) and E$_{s}$(B-V) = 0.24 $\pm$ 0.09 (0.04 $\pm$ 0.35) for the A2 (A3) component.  While the color excess of A3 is formally consistent with zero, there is a significant level of dust extinction in component A2.  From the dust and lensing-corrected H$\alpha$ flux, we calculate a SFR in the A2 component of 303 $\pm$ 84 \sfr, and in the A3 component of 67 $\pm$ 72 \sfr.  These points are also plotted in Figure 5, and we see that each component also lies near the SFR-dust trend from the Shapley et al.\ samples at z $\sim$ 2 -- 3.  The brighter component (A2) is dominating the integrated SFR of the galaxy, as well as the dust attenuation on the integrated spectrum of both components.  The metallicities are Z = 0.82 $\pm$ 0.34 (0.70 $\pm$ 0.31) Z\sol~via the N2 index, and Z = 0.40 $\pm$ 0.13 (0.38 $\pm$ 0.14) Z\sol~via the O3N2 index for the A2 (A3) component.  We thus find that while the dust and SFR properties differ dramatically between the two components, the metallicities are consistent.  This could indicate that that the current stellar population is fairly young, as the starburst in component A2 has not been occurring long enough to significantly change the metallicity (although it may have been able to significantly increase the dust content).  However, we caution that these are estimates, in that looking at Figure 2, one can see that not all of component A3 fell in the slit, thus we can only constrain the SFR of the portion of the component which was observed.

\subsection{Mass}

\subsubsection{Stellar Mass}
As mentioned in \S 3.3, our relatively high metallicity (compared to, e.g., cB58) may not be surprising given the high intrinsic luminosity of the 8 o'clock arc.  However, it may be more instructive to compare the metallicity and the mass of the arc, in the manner of Tremonti et al. (2004) and Erb et al. (2006a).  Measurements of the dynamical mass of a galaxy are possible if the emission lines widths are resolved, and if one has a measurement of the galactic radius, which is complicated due to the lensing nature of the 8 o'clock arc.  Also, the dynamics in this galaxy may not be due to orderly rotation.  However, we can still obtain a measure of the stellar mass by comparing the known photometry of the 8 o'clock arc to stellar population models.  

Using the SDSS g$^{\prime}$, r$^{\prime}$, i$^{\prime}$, z$^{\prime}$ and NIRI H and K$^{\prime}$ photometry (which are all consistent with total fluxes), we have compared the data for each component of the 8 o'clock arc to a suite of models from Bruzual and Charlot (2003).  While this process can be subject to degeneracies, specifically between the age and dust extinction, our photometry spans the age sensitive 4000 \AA~break, thus this degeneracy should be minimized.  Briefly, to estimate our stellar mass we take our measurements and compare them to a grid of models with a Salpeter initial mass function, where the stellar metallicity, age, mass and dust extinction (using the dust law from Calzetti et al. 2000) are allowed to vary.  We then find the best-fit model via $\chi^{2}$ minimization.  We allowed two different star-formation history (SFH) scenarios.  In the first, each arc component is fit to a stellar population where the SFH timescale is allowed to vary with five allowed exponential decay times, from $\tau_{SFH}$ = 10$^{5}$ -- 4 $\times$ 10$^{9}$ yr.  In the second, a two-burst (where a burst is approximated by $\tau_{SFH}$ = 10$^{5}$) stellar population model is fit to each component, where one burst is maximally old (2.4 Gyr at z = 2.73), and the other burst has an age which is allowed to vary.  This procedure is described in greater detail in Papovich, Dickinson \& Ferguson (2001) and Finkelstein et al. (2009).

The results of our fitting are shown in Figure 7.  The best-fit model for component A2 is a two-burst model with a mass of 3.0 $\times$ 10$^{11}$ M\sol, with 99\% of this mass in old stars, and 1\% in a 6 Myr population.  In this model component A2 has a mass-weighted age of 2.38 Gyr, although this is dominated by our assumption that the old burst is maximally old.  The young population could confirm our earlier speculation that component A2 is young (\S 3.4).  The best-fit model for component A3 is a single-population model with a mass of 1.2 $\times$ 10$^{11}$ M\sol, and an age of $\sim$ 1.4 Gyr.

The total stellar mass for these two components of the 8 o'clock arc is thus $\sim$ 4.2 $\times$ 10$^{11}$ M\sol.  Using our measurement of 12 + Log(O/H) = 8.58 from the N2 index, we can plot the 8 o'clock arc on the z $\sim$ 2 mass-metallicity relation (Figure 3 of Erb et al. 2006a).  We find that the arc is consistent with the z $\sim$ 2 relation, although it lies at a slightly higher mass than the highest mass point in Erb et al.  Given that the trend from low-to-high redshift is moving downwards in metallicity, we may have expected the arc to fall below the z $\sim$ 2 trend, however this is not the case. 

Finlator and Dav\'{e} (2008) study the theoretical origin of the mass-metallicity (M--Z) relation.  They find that galaxies are tightly clustered around an equilibrium value of metallicity, dependent on the stellar mass, because they tend to process gas into stars and winds at a rate similar to their gas accretion rate.  Most interesting is that they attribute the scatter in the M--Z relation to perturbing events, which bring the gas-phase metallicity out of equilibrium.  For example, a galaxy undergoing a starburst would enrich its ISM fairly quickly, much faster than infalling gas could dilute.  Over time, the low-Z infalling gas will bring the metallicity back to its equilibrium value.  However, if one was to catch a galaxy during the peak of its starburst, one could expect it to lie above the mass-metallicity relation.  The 8 o'clock arc's position on the M--Z trend may indicate that this galaxy is nearing the end of a large starburst phase.  This could happen if the SFR was higher in the past, and thus infalling gas coupled with the decreased SFR has allowed the metallicity to settle back to its equilibrium value.  It is worth noting that the position on the mass-metallicity relation is also dependent on the lensing magnification, as if the lensing magnification is different that we have assumed the derived mass will change, while the line-ratio based metallicity will not.
\begin{figure}[h]
\epsscale{1.0}
\plotone{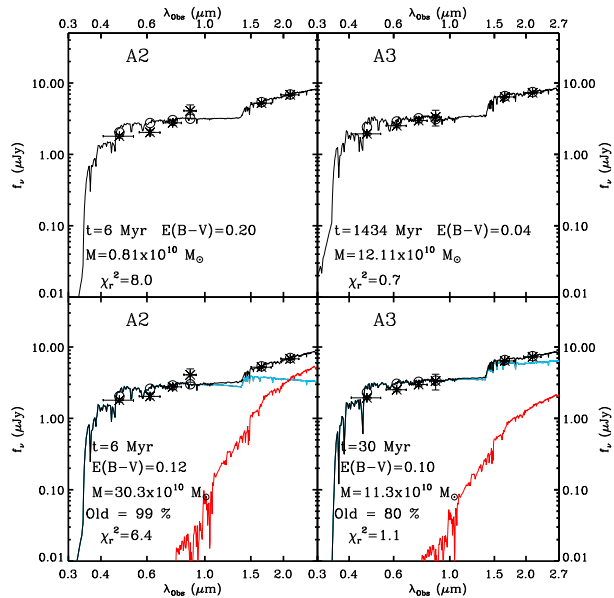}
\vspace{4mm}
\caption{Results of stellar population model fitting.  The top row shows the best single-population model to the two components, and the bottom row shows the best-fit two-burst model.  In the bottom row, the blue line is the young burst, the red line is the old burst, and the black line is the mass-weighted combination.  The percentage indicates the amount of mass in old (2.4 Gyr) stars.}
\end{figure}

\vspace{-5mm}
\subsubsection{Gas Mass}
Erb et al. (2006b) found that at z $\sim$ 2, the gas mass in a given galaxy was typically comparable to its stellar mass, thus we may expect the total dynamical mass of the 8 o'clock arc to be about twice the stellar value.  Erb et al.\ use the global Kennicutt-Schmidt law (Kennicutt 1998) with the H$\alpha$ SFR conversion we used earlier, resulting in:
\begin{equation}
\Sigma_{gas} = 1.6 \times 10^{-27}\left(\frac{\Sigma_{H\alpha}}{erg~s^{-1}~kpc^{-2}}\right)^{0.71} M_{\odot}~pc^{-2}
\end{equation}
In order to estimate the gas mass in the 8 o'clock arc, we measured the full-width half-max of the spatial extent of H$\alpha$ in the 2D spectrum, which we found to be 1.44\arcs~(after correcting for lensing), which corresponds to 11.3 kpc at z = 2.73.  This is larger than the typical radius of 6 kpc at z $\sim$ 2 from Erb et al. (2006b), but it is consistent with the 8 o'clock arc's larger stellar mass.  Correcting for extinction and lensing ($\mu$ = 8 for the combination of A2 and A3), our total H$\alpha$ flux corresponds to L$_{H\alpha}$ = 3.36 $\pm$ 0.95 $\times$ 10$^{43}$ erg s$^{-1}$.

We compute the H$\alpha$ surface density by L$_{H\alpha}$ $\times$ r$^{-2}$.  Substituting this in Equation 5, we find $\Sigma_{gas}$ = 408 $\pm$ 81 M\sol~pc$^{-2}$, and thus M$_{gas}$ = 5.25 $\pm$ 1.05 $\times$ 10$^{10}$ M\sol.  Our gas mass is thus $\sim$ 12\% of our stellar mass, contrary to Erb et al. (2006b), where the gas mass was typically $\sim$ 60\% of the stellar mass.  Given that the 8 o'clock arc is undergoing a period of heavy star formation, the low gas mass fraction could indicate that the star-forming ``fuel'' has been mostly used up.  This is consistent with our interpretation of the 8 o'clock arc's position on the mass-metallicity diagram, as discussed in \S 3.5.1.  Combining the age of the young stellar component with the current SFR, we can only account for another $\sim$ few $\times$ 10$^{9}$ M\sol~in gas, thus it is likely that the SFR was higher in the past.  Additionally, if a shock is contributing to the emission line fluxes, the wind necessary to drive that shock would also imply a strong past SFR history.  However, these results are loosely constrained, as the gas mass is derived from H$\alpha$ flux which made it through the slit, while the stellar mass is derived from the light of the entire component.  Followup estimates of the gas (from CO observations, etc.) will provide much tighter constraints on the total gas mass in the 8 o'clock arc.

\section{Conclusions}
We have presented the results of a near-infrared spectroscopic analysis of the gravitationally lensed LBG, the 8 o'clock arc.  The lensing amplification of this object has allowed us to detect rest-frame optical emission lines in the NIR with a moderate amount of observing time.  Using three detected Balmer lines, we have measured the dust extinction in the 8 o'clock arc to be A$_{5500}$ = 1.17 $\pm$ 0.36 mag.  Using our results to correct the H$\alpha$ flux for dust extinction, we have calculated the dust-corrected SFR to be 2129 $\pm$ 595 \sfr.  This falls to 266 $\pm$ 74 \sfr~when corrected for the gravitational lensing factor of $\mu$ = 4 per component (total $\mu$ = 8), putting the 8 o'clock arc in the $\sim$ 85th percentile of SFRs when compared to other samples of z $\sim$ 2--3 star-forming galaxies.

We have also measured the gas-phase metallicity using both the N2 and O3N2 indices, and we find that the 8 o'clock arc is consistent with an oxygen abundance of $\sim$ 0.8 solar.  Deriving the stellar mass of the arc via stellar population modeling, we find that this object resides on the mass-metallicity trend previously published at z $\sim$ 2.  In addition, we find that the gas mass in the arc is rather low indicating that the SFR was higher in the past.  This may point to the current burst nearing its end.

As the lens has smeared the shape of the galaxy into a partial ring, we have also been able to study two of the galactic components separately.  We found that component A2, while only marginally brighter than component A3, dominates the dust extinction in the galaxy, and is forming stars at a rate $\sim$ 4.5 times higher.  However, the metallicity of the two components is near equal, thus this could be telling us that the starburst in component A2 is fairly young, and thus has not had much time to suffuse the surrounding HII region with metals.

While other studies at comparable redshifts have derived the dust extinction and star formation rate via stellar population modeling, we have directly measured these quantities using emission lines which are difficult to detect at these redshifts, except in lensed objects.  While the next generation of 25 -- 30m class telescopes may make these observations routine at high redshifts, lensed galaxies remain our only target for detailed measurements of high-redshift galaxies for the next decade.

\acknowledgements
This work was supported by the Texas A\&M University Department of Physics, as well as by NASA though a contract issued by the Jet Propulsion Laboratory, and the California Institute of Technology under a contract with NASA.  The data presented in this paper were obtained under NOAO Program ID GN-2008A-Q-40.  ELF and JR are supported by the Spitzer Fellowship Program through a contract with JPL/Caltech/NASA.  We thank the anonymous referee for helpful comments which increased the credibility of our results.  We also thank Arjun Dey and Kate Brand for helpful conversations regarding NIRI data reduction, and Chris Fassnacht for assisting our interpretation of the gravitational lensing nature of this system.

\end{document}